\documentstyle[multicol,eqsecnum,aps,pra]{revtex}

\begin{document}

\renewcommand{\thefootnote}{\fnsymbol{footnote}}
\draft
\title{ Influence of boundaries on pattern selection  in through-flow
\footnote[1]{Dedicated to Prof. F. H. Busse on the occasion of his
60th birthday}}
\author{D.~Roth, P.~B\"{u}chel, M.~L\"{u}cke, and H.~W.~M\"{u}ller\\}
\address{Institut f\"{u}r Theoretische Physik, Universit\"{a}t
des Saarlandes, D-66041~Saarbr\"{u}cken, Germany\\}
\author{M.~Kamps\\}
\address{H\"{o}chstleistungsrechenzentrum, Forschungszentrum,
         D-52425~J\"{u}lich, Germany\\}
\author{R.~Schmitz\\}
\address{Institut f\"{u}r Festk\"{o}rperforschung, Forschungszentrum,
         D-52425~J\"{u}lich, Germany\\}
\renewcommand{\thefootnote}{\arabic{footnote}}
\renewcommand{\thefootnote}{\arabic{footnote}}
\setcounter{footnote}{0}

\maketitle

%
%

\begin{abstract}
The problem of pattern selection in {\em absolutely} unstable open 
flow systems
is investigated by considering the example of Rayleigh-B\'{e}nard
convection. The spatiotemporal structure of convection rolls 
propagating downstream in an externally imposed flow is determined 
for six different
inlet/outlet boundary conditions. Results are obtained by numerical 
simulations
of the Navier-Stokes equations and by comparison with the 
corresponding Ginzburg-Landau amplitude equation. A unique selection 
process is observed being a function
of the control parameters and the boundary conditions but independent 
of the history and the system length. The problem can be formulated in 
terms of a nonlinear eigen/boundary value problem where the frequency 
of the propagating pattern is the eigenvalue.
\end{abstract}

\pacs{47.54.+r, 47.20.Bp, 47.27.Te, 47.20.Ky}


\begin {multicols}{2}

%
%

\section{Introduction}

The spontaneous formation of patterns in  dissipative  systems  with a    
continuous energy supply is a commonly observed phenomenon
\cite{CH-RevModPhys-93}.  Classical    
examples are the generation of  rolls  in  Rayleigh-Benard    
convection (RBC)
\cite{CH-RevModPhys-93,BUSSE-78} 
and Taylor vortex flow (TVF) in the annulus between two
concentric cylinders
\cite{CH-RevModPhys-93,DIPS-Springer-81}. If  the  forcing  is    
sufficiently strong patterns bifurcate out of a  homogeneous    
base state and often show a periodic structure in space and/or in time. 
For a given set    
of parameters  and  boundary  conditions (BCs) one  frequently  observes 
a  discrete  or  continuous  spectrum  of  allowed  wave numbers and    
frequencies.  The   initial conditions finally determine which ones are 
selected.   
Usually secondary  instabilities     
limit the range of stable existence in  parameter  space.  
Recent    
numerical simulations of RBC 
\cite{MLK-Euro-89,MLK-PhysRevA-92,MLK-Nato-92} 
and TVF \cite{BLRS-96} suggest that this  degeneracy  of    
stable stationary patterns is lifted as soon as a flow through 
the system    
is imposed. These ''open  flow  systems''  exhibit  a  pattern    
selection mechanism which {\it uniquely}  fixes  the  dissipative    
structure. 

Spatially localized perturbations in open flow systems  are    
advected by the flow. Depending on whether the trailing front  of  a
growing perturbation of the basic state propagates within a linear
analysis in the downstream or 
in the upstream  direction the system is convectively    
or absolutely unstable, respectively \cite{BersBriggs}. 
The property of absolute   
or convective instability can be controlled externally by tuning  the
strength of the  imposed  flow 
\cite{Experiments,PL84,BABCOCK,TSAMERET}. However, this distinction 
was made only in the more recent experiments \cite{BABCOCK,TSAMERET}.
Pattern  formation   in   convectively   unstable    
systems requires a  {\it permanent}  source  of  disturbances  in  order  
to sustain steady  convection \cite{Deissler}.  The  spatiotemporal 
behavior of such "noise sustained structures" sensitively depends 
on the details of the perturbation source 
\cite{BABCOCK,TSAMERET}.
       
The  present    
paper, however, deals with the absolutely unstable situation
where the nonlinear structures are self sustained and stable. 
They are characterized in the longtime limit by time periodic fields 
with 
characteristic streamwise field profiles and    
wave number profiles. The frequency    
associated with the downstream motion of the pattern turns out  to  be    
constant in space and time. For a given set of  control  parameters    
and BCs,  the  pattern  in  the absolutely   unstable    
regime is independent  of initial conditions and history.     
  
In this paper we deal with RBC subject to    
an  imposed  lateral  shear  flow, i.e., with convection   rolls  
propagating   in   downstream direction.
  A  related  investigation  of  TVF    
has been published elsewhere \cite{BLRS-96}.  Different kinds of inlet
and outlet BCs are    
considered and  their  influence  upon  the  convection  structures  is    
systematically  investigated.  To  this  end  we   perform   computer   
simulations of the full hydrodynamical field equations that we refer 
to as Navier-Stokes equations (NSE) for short. 
The results are compared with analytical and numerical
solutions of the  corresponding  Ginzburg-Landau    
equation (GLE). The amplitude  equation most clearly reveals
that the selection process can be understood
as  a  nonlinear eigenvalue problem \cite{BLRS-96}: 
The frequency plays the role of    
the  eigenvalue, while the set of hydrodynamic  fields is the associated 
eigenfunction. Then the selection seems to result from requiring
the spatial variation of pattern envelope and phase to be as small
as possible under the imposed inlet/outlet BC which is analogous 
to ground state properties of the linear stationary Schr\"{o}dinger
equation.
  
The paper is organized as follows: Section II specifies  the  system,    
the BCs  for  the simulations, and  the  methods  of  investigation.
Section III recapitulates pattern selection properties in the presence 
of through-flow within the GLE. This provides a useful frame work
for  the   interpretation  of our results. 
The latter ones are  presented in  Section IV
which forms the  main part of    
our paper. A comprehensive analysis    
of  the  convection  patterns is given and a discussion of the 
influence of BCs on spatiotemporal properties of the selected 
structures as a function of through-flow rate. 
The conclusion gives a short summary of our results.

%
%

\section{The system}

We investigate properties of convection rolls in a viscous, 
incompressible fluid layer, that is heated from below in a homogeneous 
gravitational field directed downwards. In addition we impose a lateral 
through-flow. 

\subsection{Equations}

The system is described by the balance equations for mass, heat, and 
momentum in Oberbeck-Boussinesq approximation \cite{PL84,LL66,GZ76,C81}

\begin{equation}
0    =  - {\bf \mbox{\boldmath $\nabla$} \cdot u}\,, \label{Gkont}
\end{equation}

\begin{equation}
\partial_t\, T  =  - {\bf \mbox{\boldmath $\nabla$}
\cdot Q}\,\,\, ;\,\,\,
{\bf Q} = {\bf u}\, T - {\bf \mbox{\boldmath $\nabla$} }\, T\,,
\label{Gtemp}
\end{equation}
\begin{eqnarray}
\partial_t\, {\bf u} & = &  - {\bf \mbox{\boldmath $\nabla$} }\,
 ({\bf u : u}  +  p
- \sigma \, {\bf \mbox{\boldmath $\nabla :  $}\,\, u })
+ \sigma\, R\, T {\bf e}_z\,
\label{Gimp}
\end{eqnarray}
that we shall refer to as Navier Stokes equations (NSE).
Lengths are scaled by the thickness of the layer $d$, times by the 
vertical thermal diffusion time 
$d^2 / \kappa$, and the velocity field ${\bf u} = (u,\,v,\,w)$ by 
$\kappa /d$. Here $T$ denotes the deviation of the temperature from 
the mean temperature $T_0$ in the fluid measured in units of the  
temperature difference $\Delta T$ between 
the plates and $p$ is the reduced pressure. The ratio of the momentum
diffusivity $\nu$ and 
thermal diffusivity $\kappa$, i.\ e., the Prandtl number 
$\sigma = \nu / \kappa$ is the material parameter characterizing the 
fluid.  The through-flow is in positive $x$-direction, with the inlet 
at $x=0$ and the outlet at $x=L$. 
In the vertical direction the lower plate is located at $z=0$, 
whereas the upper one is at $z=1$.  

The system is characterized by two control parameters:
The Rayleigh number

\begin{equation}
Ra = \frac{\alpha g d^3  \Delta T}{\kappa \nu}
\end{equation}
measures the heating and the Reynolds number
\begin{equation}
Re = \frac{\overline{U} d}{\nu}
\end{equation}
measures the through-flow rate. 
Here $\alpha$ denotes the thermal expansion coefficient of the 
fluid, $g$ is the gravitational acceleration, and  $\overline{U}$ the
vertical average of the imposed lateral through-flow. 

\subsection{Conductive state}

For small Rayleigh numbers one observes a homogeneous basic state with a 
linear conductive temperature profile $T_{cond}$ and a parabolic 
Poiseuille profile $U(z)$ of the velocity field

\begin{eqnarray}
{\bf u}_{cond} = U(z) {\bf e}_x \,\,\, & ; & 
\,\,\, U(z) = 6 \sigma Re z (1-z) \,\,\,; \\ 
\,T_{cond} = \frac{1}{2} - z \, ,  
\label{Gdefucond}
\end{eqnarray}
which is globally stable for small thermal stress.
Note that the effective through-flow strength is given by the Peclet 
number 
$Pe=\sigma Re$. We have investigated in this work the case $\sigma=1$.
Thus, the comparison of through-flow induced effects in fluids of other 
$\sigma$ has to be based on common Peclet numbers.

\subsection{Transverse convection rolls}

At a critical  Rayleigh number $Ra_c(Re)$ 
\cite{GR68,MLK-Euro-89,MLK-PhysRevA-92} 
a laterally periodic solution
describing downstream traveling transverse convection rolls with axes
perpendicular to the through-flow direction bifurcates out of the basic 
state. For convenience we introduce the relative deviation 
\begin{equation}
\mu= \frac{Ra}{Ra_c \left(Re\right)} -1
\end{equation}
from the threshold as a new parameter measuring the externally 
imposed thermal stress.
We also use the reduced distance 
\begin{equation}
\epsilon= \frac{Ra}{Ra^0_c} -1
\end{equation}
from the threshold $Ra^0_c$ for zero through-flow such that
\begin{equation}
\mu= \frac{\epsilon}{1 + \epsilon_c\left(Re\right)}
\label{epsmu}
\end{equation}
with $\epsilon_c=Ra_c(Re)/Ra^0_c -1$.

In this work we will 
focus on patterns in which the roll axes are perpendicular to the
through-flow 
direction. They occur as the first instability in narrow convection channels
which suppress longitudinal rolls 
\cite{D-FluidMech-67,SM-FluidMech-72,PL84,LPL81,L-Thesis-83}
whereas in laterally unbounded systems the bifurcation threshold for
longitudinal rolls is lowest \cite{GR68}. 
In our numerical simulations we investigate the $x$-$z$ plane 
perpendicular to the roll axes ignoring any spatial variation along the roll 
axes. Thus we solve
the hydrodynamical field equations (\ref{Gkont}-\ref{Gimp}) for 
the lateral velocity component $u(x,z;t)$, the vertical velocity component 
$w(x,z;t)$, the pressure $p(x,z;t)$, and the temperature field $T(x,z;t)$.

\subsection{Boundary conditions}\label{BOUNDARYCONDITIONS}

The horizontal BCs are {\em always} no slip and perfectly heat conducting

\begin{eqnarray}
u=w=0 \,\,\, \mbox{at}\, z=0,1 \\
T=1/2\,\,\, \mbox{at}\, z=0, \,\,\, T=-1/2\,\,\, \mbox{at}\, z=1.
\end{eqnarray}
At the lateral boundaries we {\em always} enforce the Poiseuille 
flow profile
\begin{equation}
u(x,z;t)=U(z)\,\,\,\,\,\, \mbox{at}\,\,\, x=0,L.
\end{equation}

For the vertical velocity we have applied laterally no slip (N) 
conditions (cf.\ Table \ref{TAB1}) or
free slip (F) conditions. For the temperature field we have investigated 
the three lateral conditions that are listed in Table \ref{TAB1} and that 
are abbreviated by the symbols
$T_{cond}$, $T_0$, and $Q_0$, respectively.
We have called the latter thermally insulating since the diffusive part of 
the heat current $Q$ (\ref{Gtemp}) vanishes there. But it should be noted 
that the lateral Poiseuille 
flow transports heat advectively through the lateral boundaries. To specify 
the six possible combinations of $\{N,F\}$
conditions with $\{T_{cond}, T_0, Q_0\}$ conditions, that we have 
investigated, we use a 
two-letter coding. Thus, e.\ g., $N$-$Q_0$ implies $w=0$ and 
$\partial_x T=0$.  All numerical solutions have been obtained 
for a channel length of $L=50$.

\subsection{Methods of investigation}

To solve the time dependent 2D NSE (\ref{Gkont}-\ref{Gimp}) we used
an explicit finite difference numerical code that allows a flexible
and convenient incorporation of various BCs and that has 
been used sucessfully in related problems 
\cite{MLK-Euro-89,MLK-PhysRevA-92,BLRS-96}. 
The final-state solutions
obtained in this way represent transverse convection rolls that propagate 
downstream under a stationary $x$-dependent intensity envelope. 
These patterns oscillate in time with a global, i.\ e., spatially independent
frequency $\omega$ while wave number $k$, phase velocity $v_{ph}=\omega / k$
, and  convection intensity vary in $x$-direction. The frequency and the
spatial variation of the final pattern depend on $Ra$, $Re$, and the BCs
but not on parameter history. 

To characterize the convection structures we use the convective deviation

\begin{equation}
\delta u(x,z;t) =  u(x,z;t) - U(z)
\end{equation}
of the {\em lateral} velocity field from the Poiseuille flow as the order
parameter field. Compared with the vertical
velocity $w$ or the convective temperature profile $T - T_{cond}$
this field has some technical advantages (cf.\ Sec.\ \ref{PATTERNSELECTION}).
After having determined the global oscillation frequency $\omega$ of the
pattern we perform a temporal Fourier analysis 

\begin{equation}
\delta u(x,z;t) =  \sum_n \delta u_n(x,z) e^{-i n \omega t}.
\end{equation}
The zeroth component $\delta u_0(x,z)$ describes a stationary modification of 
the basic 
state's Poiseuille through-flow profile that reflects the presence 
of a stationary
secondary roll vortex system (cf.\ Sec.\ \ref{PATTERNSELECTION}).
The higher harmonics $u_n$ with $n\geq2$ are in general small compared 
to the fundamental mode

\begin{equation}
u_1(x,z) =  |u_1(x,z)| e^{i \varphi_1(x,z)}.
\end{equation}
We use $|u_1|$ to characterize the intensity profile of the pattern and
we characterize its wave number by $k=\partial_x \varphi_1$. Within the GLE 
approximation (Sec.\ \ref{PATTERNSELECTIONGLE}) the fields oscillate 
harmonically in time and only 
$|u_1|$ is nonzero. Thus the GLE cannot reproduce the stationary secondary
vortex flow discussed in Sec.~\ref{PATTERNSELECTION}.

%
%

\section{Pattern selection within the GLE}\label{PATTERNSELECTIONGLE}
The selection process for convection patterns in the presence of a 
lateral flow is a nonlinear eigenvalue / boundary value problem. 
This is easiest to understand within the GLE approximation \cite{BLRS-96}. 
However, one should keep in mind that GLE and NSE results differ 
systematically.
\subsection{GLE for propagating patterns}
Within the amplitude equation approximation the convection fields of the 
propagating 
pattern of transverse rolls have the form of harmonic waves, e.\ g.,
\begin{equation}
\delta u\left(x,z;t\right) = A\left(x,t\right) e^{i
\left(k_c x - \omega_c t\right)} \hat{u}\left(z\right) + c.c.
\label{HWAVE}
\end{equation}
The complex amplitude $A(x,t)$ is common to all fields and obeys the GLE
\cite{MLK-Euro-89,MLK-PhysRevA-92,MLK-Nato-92}
\begin{eqnarray}
\tau_0\left(\dot{A}+v_g A'\right) & = &
\mu\left(1+ic_0\right) A +  \xi^2_0\left(1+i c_1\right)A''
\nonumber \\
& & - \gamma\left(1+ic_2\right)\left |A\right|^2 A   \label{GLE}
\end{eqnarray}
subject to the homogeneous BC
\begin{eqnarray}
A(x,t)=0 \,\,\,\,\,\, \mbox{at} \,\,\, x=0,L.
\label{HBC} 
\end{eqnarray}
Dot and primes denote temporal and spatial derivatives. 
This BC effectively imposes on the fields in GLE approximation
the $N$-$T_{cond}$ condition: $T=T_{cond}$, $\delta u=0$, and 
$w=0$.
The other BCs listed in Table \ref{TAB1} 
are not realizable within the GLE approximation since it enforces 
one common amplitude for all fields.
However, we found that the NSE results for envelope profiles and
growth lengths obtained with the $F$-$T_{cond}$ BC compare better
with the GLE results.
The GLE ansatz for the fields implies via the critical 
functions a {\em spatially constant phase difference}
of $\pi/2$ between $\delta u$ and $w$. 
This is not the case for the NSE fields subject to the $N$-$T_{cond}$ BC, 
which show close to the boundaries a deviation from the above phase 
relation, while the $F$-$T_{cond}$ BC entails phase relations 
closer to those of the GLE.

All coefficients of the GLE, the critical numbers 
$\epsilon_c$,  $k_c$, $\omega_c$, 
and the eigenfunction $\hat{u}(z)$ have been calculated \cite{MLK-PhysRevA-92} 
as functions of  $Re$.
As a consequence of the system's invariance under the combined 
symmetry operation
$\left\{x \rightarrow -x,Re \rightarrow -Re\right\}$ 
$\epsilon_c$, $k_c$, $\tau_0$, $\xi^2_0$,
$\gamma$ are even in $Re$ while $\omega_c$, the group velocity $v_g$ and the
imaginary parts $c_0$, $c_1$, $c_2$  are odd in $Re$
\cite{MLK-PhysRevA-92,RLM-PhysRevE-93}.
For $\sigma=1$ an expansion in $Re$ yields 
$\epsilon_c \approx (Re/35.93)^2$, 
$k_c \approx 3.116[1+(Re/99.97)^2]$,
$\omega_c \approx 3.6475 Re$,
$\tau_0 \approx 0.07693[1-(Re/39.13)^2]$,
$\xi^2_0 \approx 0.148[1-(Re/35.24)^2]$, 
$v_g \approx 1.229 Re$,
$\gamma \approx 0.7027[1-(Re/332.6)^2]$,
$c_0 \approx Re/140.1$,
$c_1 \approx Re/39.98$,
$c_2 \approx Re/387$ \cite{MLK-PhysRevA-92}.
The next order terms are in each case two orders 
in $Re$ higher.
\subsection{Eigenvalue problem}
Let us consider small through-flow rates $Re$ and thermal stresses 
$\mu$ slightly above the boundary \cite{Deissler} 
\begin{equation} \mu^{c}_{conv} = \frac{\tau^2_0 v^2_g}{4 \xi^2_0
\left(1+c^2_1\right)}
\label{MUCCONV}
\end{equation}
between the absolutely and convectively unstable parameter regime.
We solved the GLE (\ref{GLE}) numerically with a Cranck-Nicholson algorithm 
with a resolution 
of $20$ grid points per unit length $d$ \cite{MLK-PhysRevA-92,BLRS-96}. 
The final state solution of
(\ref{GLE}) with the homogeneous BC (\ref{HBC}) is of the form
\begin{equation} \label{HWAVENOTRANSIENTS}
A(x,t)= a(x) e^{-i\Omega t} = 
R(x) e^{i [ \varphi \left( x \right) - \Omega t ]}.
\end{equation}
It oscillates harmonically with the global frequency 
\begin{equation}
\Omega= \omega -\omega_c
\end{equation}
under a stationary envelope $R(x)$ with a stationary wave number profile
\begin{equation}
q(x)= k(x)-k_c = \varphi'(x).
\end{equation}
Inserting the solution ansatz (\ref{HWAVENOTRANSIENTS}) into the GLE one 
obtains the nonlinear eigenvalue problem
\begin{eqnarray}
i\tau_0(-\Omega+v_g q)R+\tau_0 v_g R'  =  \mu(1+ic_0)R  \nonumber\\
+\xi^2_0(1+ic_1) (R''-q^2R+iq'R+2iqR') \nonumber \\
-\gamma(1+ic_2)R^3.
\label{EVPROBLEM}
\end{eqnarray}
The eigenvalue $\Omega$ and the associated eigenfunction 
$a(x)=R(x) e^{i \varphi \left( x \right)}$ 
describing dynamics and structure 
of the pattern are fixed by the BC (\ref{HBC}). 
We cannot make statements about the eigenvalue 
spectrum. But our numerical solutions of (\ref{GLE}) suggest that 
initial 
conditions always evolve into a pattern with a smooth amplitude 
$a(x)$ with small frequency $\Omega$. In that sense the eigenvalue 
problem 
(\ref{EVPROBLEM}) with the 
BC (\ref{HBC}) resembles ground state problems of the Schr\"{o}dinger 
equation -- increasing the spatial variation of the wave function 
(amplitude $a$)
implies higher kinetic energy (frequency $\Omega$).

%
%

\section{Pattern selection within the NSE 
for different boundary conditions}\label{PATTERNSELECTION}
In this section we investigate how structure and dynamics of patterns 
selected in different through-flows are influenced by the lateral 
BC. We present numerical simulations of the full field 
equations in comparison with numerical solutions of the GLE. Results are
often presented as functions of the scaled group velocity 
\cite{MLK-Euro-89,BLRS-96}.
\begin{equation}
V_g = \frac{\tau_0 v_g}{\xi_0 \sqrt{\left(1+c^2_1\right)\mu}} = 
2 \sqrt{\frac{\mu^c_{conv}}{\mu}}
\label{Vg}
\end{equation}
which scales as $Re/\sqrt{\mu}$ for small $Re$ 
(cf.~ Sec.~\ref{PATTERNSELECTIONGLE}).
With this scaling the boundary between absolute and convective instability
lies at $V_g=2$ and thus allows direct comparisons for different 
$\mu$-$Re$ combinations. We have varied the Reynolds number at a fixed 
$\varepsilon=0.215$. Thus we scan the $Re$-$\mu$ control parameter plane 
according
to (\ref{epsmu}) along the path 
$\mu=0.215/\left[1+\varepsilon_c(Re)\right]$ that is slightly 
curved and that hits the borderline $\mu_{conv}^c$ (\ref{MUCCONV}) 
of the absolutely 
unstable regime at $Re \simeq 3.8$. Along this path $V_g$ varies for 
$\sigma=1$ as $V_g \approx 0.53 Re + O(Re^3)$. 

\subsection{Selected frequency eigenvalues}

In Fig.~\ref{figa} we show NSE frequency eigenvalues (symbols) as 
functions of the scaled group velocity $V_g$ (\ref{Vg}) for the six
BCs of Table \ref{TAB1} together with the GLE eigenvalue
(full line) for $N$-$T_{cond}$ conditions.
Approaching the border to the convective instability at 
$V_g=2$ the growth length of the patterns from the inlet increases and the 
dependence of $\omega$ on the BC gets weak and all $\omega$ values 
come close together, whereas for small $V_g$ much stronger differences 
can be 
observed. The selected frequencies are predominantly influenced by the 
inlet BC.
A change of the outlet BC leads to $\omega$-variations which are 
$10^{-3}$ to $10^{-5}$ times smaller than those induced by changing
the inlet BC.
This insensitivity of $\omega$ to the outlet BC reflects the 
 different amplifying properties 
of perturbations in upstream or downstream direction. 
However, the intensity envelopes (Sec.~\ref{PATTERNSELECTION}B) and the 
wave 
number profiles (Sec.~\ref{PATTERNSELECTION}D) are more strongly 
influenced by the outlet 
BC, in particular in its vicinity.

From  all BCs $N$-$T_{cond}$ leads to the largest and $F$-$Q_0$ to 
the smallest frequencies. In general a change in the inlet BC from no slip
to free slip causes $\omega$ to decrease.  
For $T_{cond}$  and $T_0$ conditions this change  is 
small, whereas it becomes significant for the thermally insulating cases $Q_0$.  
This can be explained by the different phase pinning properties of the 
different BCs. In the $N$-$T_{cond}$ case that enforces the basic state 
at in- and outlet the temporal fourier amplitudes of 
{\em all} convective fields drop to zero, leaving the phase 
indetermined and free to move.
On the other hand, enforcing a von Neumann BC for one of the fields yields 
a nonvanishing  
field amplitude, a phase pinning effect, and therefore a reduction of the 
phase velocity. 
In the case of $F$-$Q_0$ the phase pinning is strongest, because there 
$\partial_x w$ 
as well as $\partial_x T$ vanish, leaving a free-phase condition only for 
$u$. 

Both $T_0$ conditions induce a {\em large} stationary "boundary vortex" flow 
in the vicinity of the boundaries, that can 
be seen directly via the zeroth Fourier modes of the fields, 
e.\ g., $\delta u_0$
and indirectly via
the lateral variation of, e.g., $|u_1|$ (cf.~Sec.~\ref{PATTERNSELECTION}B and 
Fig.~\ref{figb}). It is quite similar to 
the stationary Ekman vortex structures that
appear in the Taylor Couette system with through-flow as a result 
of a vanishing
azimuthal velocity at inlet and outlet \cite{BLRS-96}. 
Besides that the lateral variation of $|u_1|$ induces for all BCs weaker 
stationary "flank vortices" (cf.~Sec.~\ref{PATTERNSELECTION}B and 
Fig.~\ref{figb})
localized in the region of largest lateral growth of $|u_1|$.
The stationary vortex structures exert an additional
frictional force on the phase of the propagating vortices 
that reduces their phase 
velocity, when the fields $\delta u_0$ and $u_1$ spatially overlap.
The frequency
reduction roughly depends on the $\delta u_0$-amplitudes of the 
stationary secondary vortex structures relative to the size of the 
propagating vortex amplitudes, say $|u_1|$. An analogous argument
was given by Cross et al. \cite{CDHS-FluidMech-83}  for the wave number 
selection at $Re=0$.
In this way one can understand {\it (i)} That the $\omega$-reduction for
$T_{cond}$ and $Q_0$ boundaries saturates with decreasing $V_g$: For these 
BCs the magnitude of the $\delta u_0$-amplitudes of the "flank vortices" 
approaches near the boundaries the size 
of $|u_1|$ from below when the $|u_1|$-profile of the propagating structure
comes closer and closer towards the inlet for $V_g \rightarrow 0$.
{\it (ii)} That $\omega$ decreases for small $V_g$ dramatically for 
$T_0$-boundaries: They enforce stationary "boundary vortex" amplitudes 
$\delta u_0 > |u_1|$ such that $\delta u_0/|u_1|$
further increases when $V_g \rightarrow 0$.
As an aside we mention that the decrease of $\omega / \omega_c -1$ 
at small through-flow in the
$T_0$ case differs from the otherwise very similar 
results for propagating vortex patterns in the rotating Couette system with
Ekman vortex enforcing inlet/outlet conditions. There an increase 
of $k/k_c-1$,
i.\ e., of $\omega/\omega_c-1$ has been observed for small $Re$ 
\cite[Fig.~6]{BLRS-96}.

The GLE frequency eigenvalues (full curve in Fig.~\ref{figa}) deviate 
in a systematic manner from the NSE ones (open circles
for the common $N$-$T_{cond}$ BC). This has also been found for downstream 
propagating
vortices in the Taylor Couette system \cite{BLRS-96}. The reason for this 
systematic deviation 
seems to be here as for propagating Taylor vortices
the different dispersion relations of NSE and GLE.  
See Ref.~\cite{BLRS-96} for a detailed discussion.
\subsection{Lateral pattern profiles}
In Fig.~\ref{figb} we present the lateral variations of $|u_1|$ 
and $\delta u_0$ 
at $z=0.225$ for two different $Re$ values. The bulk region 
$7<x<45$ is cut out, because there the Fourier modes are almost constant. 
Since $\delta u$
is zero at in- and outlet for all our BCs this field allows
best to elucidate the changes induced by different BCs. 
 The lateral velocity $\delta u$ has
a further advantage for the analysis of the BC-dependence of 
convection patterns 
since $\delta u$ is more sensitive to lateral BCs than, say, $w$: The 
lateral deformations in the flanks of the profiles of 
$\delta u$ are more pronounced than for $w$. In our earlier
work \cite{MLK-Euro-89,MLK-PhysRevA-92,MLK-Nato-92} where $w$ profiles 
were analyzed we detected only minor
differences between NSE and GLE profiles where these deformations are 
completely absent.

The deformations in the lateral field profiles of, say $|u_1|$, 
are caused by the 
nonlinear couplings to a stationary vortex structure located
in the flank regions where the lateral variation of the intensity 
$|u_1|$ of the 
propagating pattern is large. Typically the secondary flow amplitude 
$\delta u_0$ 
of the "flank vortices" reaches a strength of about 2\% of the bulk
$|u_1|$.
The "flank vortex" flow increases when changing the BCs according to 
$N \rightarrow F$ and/or $T_{cond} \rightarrow T_0 \rightarrow Q_0$. The 
"flank vortices" associated with the inlet facing flanks of $|u_1|$
become weaker with increasing $V_g$ since then the $|u_1|$-profile becomes 
smoother as its growth length increases (see Fig.~\ref{figb}d for $Re=8/3$). 

Besides this $|u_1|$-induced
stationary "flank vortex" flow, there is for the $T_0$ BC a 
boundary induced stationary "boundary vortex" 
structure of {\em large} amplitude $\delta u_0 \approx |u_1|$
which is very similar to the stationary Ekman vortex structure in the
analogous Taylor Couette system \cite{BLRS-96}. This "boundary vortex"
structure, induced 
by the $T_0$ BC, is always located near the lateral boundaries irrespective 
where the $|u_1|$ flanks are located and its intensity falls off towards 
the bulk
-- see $\delta u_0$ in Fig.~\ref{figb}d for $Re=8/3$.

For a more quantitative comparison of the selected amplitude's profiles 
we have determined the 
characteristic growth lengths from inlet and outlet where 
$|u_1|$ reaches half of its bulk value. 
To that end we have performed a least square fit to $|u_1|$ with the function
$(1-e^{a x})/(b+e^{a x})$
with fit parameters $a$ and  $b$
in order to obtain also in the presence of amplitude deformations 
comparable results. This fit function yields good results 
except directly in the inlet region.
In Fig.~\ref{figc} 
we show the scaled growth lengths 
\cite{MLK-Euro-89,MLK-PhysRevA-92,MLK-Nato-92}
\begin{equation}
L=\sqrt{\mu} \, l/\xi_0
\label{L}
\end{equation}
of $|u_1|$ as a function of $V_g$.
Note that inlet (outlet) growth lengths are shown for positive 
(negative) $V_g$.
$L$ has to be identical at in- and outlet for  $V_g=0$.
In addition $L$ seems to have there a continuous first derivative for 
$T_{cond}$ and $Q_0$ conditions but not for $T_0$. For the latter 
BC $L$ strongly
increases as $V_g$ approaches zero since the stationary "boundary vortex"
induced by the $T_0$ condition strengthens with decreasing $|V_g|$ 
relative to $|u_1|$
and therefore pushes the oscillating pattern into the bulk.
Changing the BCs along the sequence 
$F$-$T_0 \rightarrow N$-$T_0 \rightarrow N$-$T_{cond} 
\rightarrow F$-$T_{cond} \rightarrow N$-$Q_0 \rightarrow F$-$Q_0 $ 
decreases $L$
successively.
For both $T_{cond}$ conditions GLE and NSE agree quite well with
deviations being smaller for $F$-$T_{cond}$ BC than for 
$N$-$T_{cond}$ BC (see Sec.~\ref{PATTERNSELECTIONGLE}A).
\subsection{Phase relations in the bulk}

Let us consider first the dispersion relation between the frequency 
$\omega$, i.e., 
the temporal phase gradient and the spatial gradient $k$ in the 
bulk part of the system.
The GLE (\ref{EVPROBLEM}) predicts the relation 
\begin{equation}
\tau_0\Omega= (c_2-c_0)\mu + \tau_0 v_g q_b +  (c_1-c_2)\xi^2_0 q^2_b
\label{OMQBULK}
\end{equation}
between selected frequency $\Omega=\omega-\omega_c$, wave number
$q_b=k_b-k_c$, and modulus $R^2_b=(\mu-\xi^2_0 q^2_b)/\gamma$
for  bulk patterns with  $R'_b = R''_b = q'_b = 0$.
Fig.~\ref{figd}a shows the GLE result (\ref{OMQBULK}) (full line) together 
with the NSE results
(symbols) obtained for the six BCs for all Reynolds numbers sufficiently
large to establish a bulk region with spatially constant wave number.
Since the bulk GLE wave numbers are very close to the critical one 
the quadratic
contribution $q_b^2$ to $\Omega$ and to $R_b$  is for all 
through-flow rates
$0<V_g<2$ 
very small so that $\Omega$ is practically linear in $k$  (Fig.~\ref{figd}a).
That also holds for the NSE case. Here, however, the spread in the bulk wave
numbers around $k_c$ is considerable larger depending on the lateral 
BC and $Re$.
Within the resolution of Fig.~4a the bulk dispersion relation for 
propagating roll
vortices resulting from GLE or NSE reads
\begin{equation}
\omega \simeq \omega_c+v_g(k-k_c)
\label{OMQBULKapprox}
\end{equation}
irrespective of the BC. The slight deviation between the thin dotted 
curve in Fig.~\ref{figd}a
that represents $\omega/\omega_c -1 \simeq (k-k_c)v_g/\omega_c $ according to
 (\ref{OMQBULKapprox}) and the GLE result (full line) comes from 
 the contribution
$(c_2-c_0)\mu/(\omega_c\tau_0)$ to the GLE dispersion relation (\ref{OMQBULK}).

In Fig.~4b we show the relation between bulk intensity $|u_1|^2$ of 
the propagating
pattern and bulk wave number. Again the symbols (full line) denote the 
NSE (GLE) 
results for several $Re$ and BCs. To determine $|u_1|^2$ for the GLE we 
have used
(\ref{HWAVE}) with the properly normalized linear eigenfunction $\hat{u}(z)$ 
\cite{CJ-Privat} evaluated at $Re=0$.
The relations between bulk pattern amplitude and bulk wave number 
resulting from NSE 
and GLE are obviously quite different. 
The peculiar shape of the full curve in Fig.~\ref{figd}b which seems to be at 
first sight at odds with the GLE 
formula $R^2_b = \mu - \xi^2_0 q^2_b$ is caused
by two facts:
(i) Due to the pattern selection process the selected wave number in the bulk region is
uniquely determined by the control parameters $\mu$, $Re$. For a given $\mu$ one 
obtains for $Re=0$ the critical wave number \cite{CDHS-FluidMech-83}. Then, 
with increasing $Re$ the wave number first slightly increases and later on 
it decreases below the critical one to the value
\begin{equation}
\xi_0 q_b=
-\frac{\sqrt{1+c^2_1} - \sqrt{1+c^2_2} }{c_1-c_2}
\sqrt{\mu^c_{conv}}
\end{equation}
at the border between absolute and convective instability \cite{BLRS-96}.
(ii) The contribution from $\xi^2_0 q^2_b$ to $R^2_b$ is very small since 
$|q_b| \lesssim 0.05$. Therefore the variation of $R^2_b$ and thus of
$|u_1|^2$ is mainly
caused by monotonously decreasing $\mu= 0.215/[1+(Re/35.93)^2]$ while
increasing $Re$ from $Re=0$ to $Re \simeq 3.8$ where $V_g=2$.

\subsection{Wave number profiles}
Having investigated  the frequency eigenvalue
and the modulus profile $|u_1|$ of the associated eigenfunction we now 
address its phase structure. 
The frequency eigenvalue and the lateral profile of the phase gradient
$k(x)=\partial_x \varphi_1(x)$ seem to be closely tied together. And the 
latter is largely slaved by the envelope profile $|u_1|$ which itself is most 
strongly influenced by the BCs.

Fig.~\ref{fige} shows  selected  modulus  profiles  of  $|u_1|$  together  
with 
corresponding wave number profiles $k(x)$  for  a  small  through-flow 
with $Re=1/6$. The inlet boundary  conditions  were  $N$-$T_{cond}$  or
$F$-$Q_0$ and  the  outlet  conditions  were  always  $N$-$T_{cond}$.  The 
frequencies  for  these  two  inlet BCs differ 
significantly (c.f. Fig.~1) and also the wave number  profiles  $k(x)$ 
are quite different.  Large  deformations  in  the  pattern  amplitude 
$|u_1|$ cause strong deformations of the  phase  gradient  $k(x)$.  So 
both, the temporal derivative $\omega$ and the spatial gradient $k(x)$ 
of the phase of the propagating pattern are influenced  by  the  inlet 
BC. Note also that the (boundary induced) deformations of the  pattern 
profile $|u_1|$ correlate with corresponding variations of $k(x)$. The 
GLE, on the other hand, yields a wave number profile (open circles  in 
Fig.~5b) that is very close to $k_c$ and that is  practically  constant 
(c.f. \cite[Fig.~4b]{BLRS-96} for more details) in space while  the  
NSE  wave
numbers show significantly larger variations. 

In the inset of Fig.~5 we show $k(x)$ in  the  bulk  region  $10<x<45$ 
where $|u_1|$ is almost constant. For the phase pinning $F$-$Q_0$ BC the 
wave number  does  not  saturate  into  a  bulk  part  for  the  small 
through-flow of $Re=1/6$. Instead $k(x)$ shows an  exponential  growth 
(full line in the inset of Fig.~5)  in  the  region  of  bulk  pattern 
amplitude from small $k$ in the upstream portion to large $k$  in  the 
downstream region:  For  sufficiently  small  through-flow  the  phase 
pinning forces at the inlet are strong enough to cause  an  elongation 
of the pattern, i.e., a decrease in $k$ near the inlet that heals  out 
exponentially in downstream direction. Increasing $Re$  or  decreasing 
the inlet phase pinning with other BCs reduces the pattern elongation, 
i.e., the wave number gradients. On the other hand,  decreasing  $Re$ 
increases the gradient of $k(x)$ and eventually  even  suppresses  the 
phase propagation altogether. This latter  transition  at  very  small 
$Re$  depends on  the  system 
size --- c.f. Ref.\cite{MLK-PhysRevA-92} for a related detailed 
investigation.

However, when the parameters are such that a true bulk part of the 
pattern is
established in which amplitude {\em and} wave number are constant then
the selected frequency and lateral pattern profile seem to be independent
of the system size. This always occurs in sufficiently large systems
when the through-flow is sufficiently strong.

%
%

\section{Conclusion}

We have investigated convection roll structures in
systems of finite length that propagate downstream in an externally
applied lateral flow with roll axes perpendicular to the flow
direction subject to six different inlet/outlet BCs.  Within the
absolutely unstable parameter regime a unique pattern selection is
observed. The selected convection patterns are independent of
parameter history, initial condition, and system length provided the
latter is large enough to allow for a saturated bulk region with
homogeneous pattern amplitude and wave number. But they depend on the
lateral BC. The outlet BC influences the convection structures only
locally in the vicinity of the outlet while the globally constant
frequency and the lateral profiles of pattern amplitude and wave
number and their bulk values are basically determined by the inlet BC.
The selected frequency $\omega$ of the pattern oscillation is an
eigenvalue of a nonlinear eigenvalue problem with the set of time periodic
convection fields, $f(x,z,t) = f(x,z,t+2\pi/\omega)$, being the 
associated
eigenfunction.  The analysis of the appropriate GLE approximation
suggests that the eigenfunction associated to the selected
eigenfrequency varies as smoothly and as little as possible under the
imposed BC. This property is similar to the ground state behavior of
the linear stationary quantum mechanical Schr\"{o}dinger equation.
 
Different lateral BCs entail different frequencies and eigenfunctions,
i.e., lateral pattern profiles. For example, inlet conditions that
exert a phase pinning force on one or more of the fields of traveling
convection rolls reduce the frequency. The spatial growth behavior and
the pattern profile of the convective structures in the region between
inlet and bulk saturated pattern is strongly influenced by the inlet
BC and can be quite intricate. While the frequencies and bulk wave
numbers obtained for various BCs and through-flow rates well inside
the absolutely unstable regime differ substantially they all fall onto
a linear dispersion curve.  Finally, when approaching with
increasing $Re$ the border between absolute and convective instability
the eigenfrequencies and also the spatial pattern profiles belonging
to different BCs approach each other --- with increasing growth length
of the pattern from the inlet the influence of the latter on
spatiotemporal properties of the convection pattern becomes weaker.

%
%

\acknowledgments

This work was supported by the Stiftung Volkswagenwerk.

%
%

%
%

\narrowtext

\begin{table}
\caption{Lateral boundary conditions at $x=0$ and/or $x=L$.}

\begin{minipage} [b]{8.5cm}

\begin{tabular}{|c|c|c|}
field & type & symbol \\ \hline
$w(x,z;t)=0$ & no slip & N \\
$\partial_x w(x,z;t)=0$ & free slip & F \\
$T(x,z;t)=T_{cond}(z)$ &  conductive profile & $T_{cond}$ \\
$T(x,z;t)=0$ & mean temperature & $T_0$ \\
$\partial_x T(x,z;t)=0$ & insulating & $Q_0$ \\
\end{tabular}
\end{minipage}

\label{TAB1}
\end{table}

%
%


\begin{figure} \caption[]
{Selected oscillation frequency of traveling convection rolls obtained
from the NSE for different lateral BCs (symbols) and from the GLE for 
$N$-$T_{cond}$ BC (full line) versus the scaled group velocity $V_g$
(\ref{Vg}). Dashed lines are guides to the eye. Parameters are 
$\epsilon=0.215$ and $\sigma=1$.}
\label{figa}
\end{figure}


\begin{figure} \caption[]
{Lateral profiles of convection structures for different BCs.
Temporal Fouriermodes $|u_1|$ and $\delta u_0$ reduced
by the bulk value of $|u_1|$ at $x=30$ are shown at $z=0.225$
close to the 
inlet ($x=0$) and outlet ($x=50$) boundaries. For clarity 
$\delta u_0$ is suppressed in the outlet region $47<x<50$.
Lines refer to NSE results and open circles to the GLE
subject to $N$-$T_{cond}$ conditions. Parameters are 
$\epsilon=0.215$ and $\sigma=1$.}
\label{figb} 
\end{figure}


\begin{figure} \caption[]
{Scaled growth lengths $L$ of propagating convection patterns versus 
scaled
group velocity $V_g$. Growth distances where $|u_1|$ reaches half
its bulk value from inlet (outlet) are shown for positive (negative) 
$V_g$. Symbols refer to NSE results and line to GLE. Parameters are 
$\epsilon=0.215$ and $\sigma=1$.}
\label{figc}
\end{figure}


\begin{figure} \caption[]
{ (a) Frequency eigenvalues $\omega/\omega_c-1$ and (b) bulk flow
intensities $|u_1|^2$ versus selected bulk wave numbers $k-k_c$ of
propagating roll patterns are shown for several Reynolds numbers
covering the whole absolutely unstable regime $0<V_g<2$.
Symbols (lines) refer to results obtained from the NSE with
different BCs (GLE with $N$-$T_{cond}$ BC). The thin dotted curve in (a)
shows $(k-k_c)v_g/\omega_c$ as discussed in Sec.~IV C. Parameters are 
$\epsilon=0.215$ and $\sigma=1$.}
\label{figd}
\end{figure}


\begin{figure} \caption[]
{(a) Amplitude profiles and (b) wave number profiles close to inlet and
outlet. The temporal Fourier mode $|u_1|$ of the traveling 
convection pattern was reduced by the bulk value at $x=30$. The 
inset shows 
the lateral variation of the wave number in the center region of
the system. The outlet BC was $N$-$T_{cond}$, the inlet BCs are
indicated in the inset. Parameters are 
$\epsilon=0.215$ and $\sigma=1$.}
\label{fige}
\end{figure}

\end   {multicols}
\end{document}